# STOCHASTIC PARTICLE ACCELERATION
# NEAR ACCRETING BLACK HOLES

Charles D. Dermer[1], James A. Miller[2], and Hui Li[3]

## ABSTRACT

We consider the stochastic acceleration of particles which results from resonant interactions with plasma waves in black hole magnetospheres. We calculate acceleration rates and escape time scales for protons and electrons resonating with Alfvén waves, and for electrons resonating with whistlers. Assuming either a Kolmogorov or Kraichnan wave spectrum, accretion at the Eddington limit, magnetic field strengths near equipartition, and turbulence energy densities $\sim 10\%$ of the total magnetic field energy density, we find that Alfvén waves accelerate protons to Lorentz factors $\lesssim 10^4$–$10^6$ before they escape from the system. Acceleration of electrons by fast mode and whistler waves can produce a nonthermal population of relativistic electons whose maximum energy is determined by a competition with radiation losses.

Particle energization and outflow is not possible at lower accretion rates, magnetic field strengths, or turbulence levels due to dominant Coulomb losses. Increases in the accretion luminosity relative to the Eddington luminosity can trigger particle acceleration out of the thermal background, and this mechanism could account for the differences between radio-quiet and radio-loud active galactic nuclei. Observations of outflowing radio-emitting components following transient X-ray events in galactic X-ray novae and gamma-ray flares in blazars are in accord with this scenario.

*Subject headings:* galaxies: active — galaxies: Seyfert — gamma-rays: galaxies — radiation processes — particle acceleration

[1] E. O. Hulburt Center for Space Research, Code 7653, Naval Research Laboratory, Washington, DC 20375-5352; E-mail: dermer@osse.nrl.navy.mil

[2] Department of Physics, University of Alabama in Huntsville, Huntsville, AL 35899; E-mail: millerj@cspar.uah.edu

[3] Department of Space Physics and Astronomy, Rice University, Houston, TX 77251. Present address: NIS-2, MS D436, Los Alamos National Laboratory, Los Alamos, NM 87545; E-mail: li@ssctx1.lanl.gov

astro-ph/9508069    16 Aug 1995

## 1. Introduction

The high efficiency of energy generation inferred from radio observations of quasars (Lynden-Bell 1969) and X-ray observations of Seyfert AGNs (Fabian 1979) is apparently achieved only by the gravitational conversion of the rest mass energy of accreting matter onto supermassive black holes. Evidence for the acceleration of particles to high energies by a central engine is inferred from observations of apparent superluminal motion in flat spectrum, core-dominated radio sources (Zensus & Pearson 1987). This phenomenon is widely attributed to the ejection of relativistic bulk plasma from the nuclei of active galaxies (Blandford & Rees 1978), and accounts for the existence of large scale radio jets and lobes at large distances from the central regions of radio galaxies (e.g., Begelman, Blandford, & Rees 1984). Reports of radio jets (Mirabel et al. 1992) and superluminal motion (Mirabel & Rodríguez 1994) from galactic black hole candidate X-ray sources indicate that similar processes are operating in these sources. Observations (Fichtel et al. 1994; Johnson et al. 1994; McNaron-Brown et al. 1995) of luminous, rapidly variable high-energy radiation from active galactic nuclei (AGNs) with the *Compton Gamma Ray Observatory* show directly that particles are accelerated to high energies in a compact environment (see Dermer & Gehrels 1995 for interpretation).

The mechanisms which transform the gravitational potential energy of the infalling matter into nonthermal particle energy in galactic black hole candidates and AGNs are not conclusively identified, although several have been proposed (see, e.g., Blandford 1990; Eilek & Hughes 1991 for a review of acceleration processes). These include direct acceleration by static electric fields (resulting from, for example, magnetic reconnection), shock acceleration, and energy extraction from the rotational energy of Kerr black holes (Blandford & Znajek 1977). The dominant acceleration mechanism(s) operating in the black hole environment can only be determined, of course, by a comparison of model predictions with observations.

In this paper, we consider stochastic particle acceleration through resonant interactions with plasma waves that populate the magnetosphere surrounding an accreting black hole. Stochastic acceleration has been successfully applied to the problem of ion and electron energization in solar flares (e.g., Melrose 1974; Barbosa 1979; Eichler 1979; Ramaty 1979; Dröge & Schlickeiser 1986; Miller & Ramaty 1989; Miller & Viñas 1993; Mazur et al. 1992), and is capable of accounting for a wide range of both neutral and charged particle emissions (see also reviews by Ramaty & Murphy 1987; Miller et al. 1995). It is also a component in diffusive shock acceleration, since pitch-angle scattering (which is necessary for multiple shock crossings) is accompanied by diffusion in momentum space, which in turn yields a net systematic energy gain; however, stochastic energization will dominate the first-order shock process only in certain parameter regimes (Schlickeiser, Campeanu, & Lerche 1993). Although stochastic acceleration has been applied to particle energization in the lobes of radio galaxies (e.g., Lacombe 1977; Achterberg 1979; Eilek 1979; Bicknell & Melrose 1982; Eilek & Henriksen 1984), its application to the central regions of AGNs has only recently been considered (Henri & Pelletier 1991), but not in detail. Here we systematically investigate the plasma processes responsible for stochastic particle acceleration along with the energy-loss processes which impede particle energization. Application of this work to ion abundance enhancements in galactic black holes is considered separately (Miller & Dermer 1995).

In § 2.1, we describe the geometry of our model for stochastic particle acceleration in a black hole magnetosphere, and establish standard parameters for the magnetic field strength using equipartition arguments. In § 2.2, we present the physics of stochastic acceleration by resonant interactions with parallel transverse plasma waves. Employing the quasilinear approximation, diffusion coefficients for protons and electrons interacting with Alfvén waves, and electrons interacting with whistlers, are calculated and given in §§ 2.2 and 2.3, respectively, along with the associated energy-gain rates and time scales for particles to diffuse out of the system. The competition between acceleration and energy losses for protons and electrons are considered in §§ 3.1 and 3.2, respectively, where we find that acceleration of protons to high energies can occur for standard parameters, although proton acceleration can be prevented at lower accretion rates, magnetic field strengths, or weaker levels of the turbulent energy density. Application of these results to AGN and galactic black hole observations is considered in § 4, where we propose a triggering mechanism to account for the difference between radio-loud and radio-quiet AGNs. Calculations of the detailed radiative signatures of this system are, however, deferred to a later paper. We summarize in § 5. In Appendix A, we give an intuitive derivation of the stochastic diffusion coefficients, and in Appendix B, we derive a formula for the electron acceleration time scale which bridges the low-frequency fast mode and the whistler regime.

## 2. Stochastic Particle Acceleration in Black Hole Magnetospheres

### 2.1. Model Geometry

Our model geometry is shown in Figure 1. An accretion disk forms around a supermassive black hole of mass $10^8 M_8$ Solar masses. The accreting plasma is assumed to support a large scale axisymmetric magnetic field configuration due to the generation of currents (see, e.g., Begelman et al. 1984; Li, Chiueh, & Begelman 1992 and references therein). In this paper, we average the plasma and magnetic field properties over a spherical volume in the black hole magnetosphere of radius $R$, which we call the corona. We work in units of gravitational radii

$$R_g \equiv \frac{GM}{c^2} = 1.48 \times 10^{13} M_8 \text{ cm} , \qquad (1)$$

and we scale the size of the corona to 100 gravitational radii, so that $R = 10^2 R_2 R_g$. The coronal thermal particle number density $n_p$ can be written in terms of the Thomson depth $\tau_p = n_p \sigma_T R$, where $\sigma_T$ is the Thomson cross section, so that

$$n_p = \frac{\tau_p}{\sigma_T R} = 1.02 \times 10^9 \frac{\tau_p}{R_2 M_8} \text{ cm}^{-3}, \tag{2}$$

We scale the luminosity $L$ liberated within radius $R$ by the relation $L = l_{Edd}L_{Edd}$, where the Eddington luminosity $L_{Edd} = 1.26 \times 10^{46}M_8$ ergs s$^{-1}$. Also, recalling the definition $\ell = L\sigma_T/4\pi R m_e c^3 = L/[R(4.64 \times 10^{29}\text{ergs cm}^{-1}\,\text{s}^{-1})]$ for the compactness parameter, we see that $\ell = 18.4\, l_{Edd}/R_2$.

To assign a value for the average coronal magnetic field strength $B$, we assume that there is rough equipartition of the magnetic field energy density $U_B \equiv B^2/8\pi$ with the radiation energy density, and therefore the accretion energy density. The equipartition field strength $B_{eq}$ is given by the expression

$$\frac{B_{eq}^2}{8\pi} \simeq \frac{L}{4\pi R^2 c} \ . \tag{3}$$

implying that

$$B[\text{Gauss}] = b B_{eq} = 619 \ \frac{b l_{Edd}^{1/2}}{R_2 M_8^{1/2}} \ , \tag{4}$$

where $b \lesssim 1$.

Assuming a negligible contribution from elements heavier than H, the Alfvén velocity $v_A = B/(4\pi n_p m_p)^{1/2}$, where $m_p$ is the proton mass. From equations (2) and (4), $v_A$ in units of $c$ is given by

$$\beta_A = 0.14b \ \left(\frac{l_{Edd}}{\tau_p R_2}\right)^{1/2} \ . \tag{5}$$

The dynamical time scale of the system is given by

$$t_{dyn}[\text{s}] \equiv R/c = 4.9 \times 10^4 R_2 M_8 \ , \tag{6}$$

and the Thomson energy-loss time scale for an electron is given by

$$t_T \equiv \left|\frac{\dot{\gamma}_c}{(\gamma - 1)}\right|^{-1} = \frac{3t_{dyn}}{4\ell} \ \frac{\gamma - 1}{p^2} \ , \tag{7}$$

where $-\dot{\gamma} = 4c\sigma_T U_{ph} p^2/3$ is the Thomson energy-loss rate, $p = \beta\gamma$ is the dimensionless particle momentum, and $\gamma = (1 - \beta^2)^{-1/2}$ is the particle Lorentz factor. Note that $t_T \ll t_{dyn}$ in a compact ($\ell \gg 1$) system, if most scattering takes place in the Thomson regime.

## 2.2.   Stochastic Acceleration by Alfvén Turbulence

Waves in a magnetized plasma have, in general, an electric field component transverse to the ambient magnetic field $\vec{B}_0$, as well a component parallel to $\vec{B}_0$. These two components can strongly affect charged-particle motion through resonant interactions (e.g., Karimabadi, Krauss-Varban, & Terasawa 1992 and references therein). A resonant interaction between a particle and the transverse electric field of a wave occurs when the Doppler-shifted wave frequency is a multiple of the particle cyclotron frequency in the particle guiding-center frame, and the sense of rotation of the transverse electric field and the particle gyrational motion is the same. In this case, depending upon the initial relative phase of the particle and wave, the particle will see either an accelerating or decelerating electric field along its transverse direction of motion over a substantial fraction of a cyclotron period, resulting in a relatively large energy gain or loss, respectively. (If the resonance condition is not satisfied, then the time over which a particle can be influenced by a wave is much shorter, and the effect on the particle is negligible.) This type of resonance primarily changes the perpendicular particle energy. On the other hand, resonance between a particle and the parallel electric-field occurs when the parallel speed of the particle approximately equals the parallel phase speed of the wave. In this case, depending upon whether the parallel particle speed is slightly less than or greater than the parallel phase speed, the particle will become trapped in a potential trough and either be accelerated up to phase speed or decelerated down to this speed, respectively. This resonance alters only the parallel energy of the particle.

The resonance condition can therefore be expressed as

$$\omega - k_\parallel v_\parallel - \frac{l\Omega}{\gamma} = 0 , \tag{8}$$

where $\omega$ and $k_\parallel$ are the frequency and parallel wavenumber of the wave; and $v_\parallel$, $\gamma$, and $\Omega = |q|B/mc$ are the parallel speed, Lorentz factor, and nonrelativistic gyrofrequency of the particle, which has charge $q$ and mass $m$. The harmonic number $l$ is equal to 0 for resonance with the parallel electric field, but equal to $\pm 1, \pm 2, \ldots$ for resonance with the transverse electric field. The $l = +1$ case is usually called the normal Doppler resonance, and is relevant when the handedness of the wave transverse electric field is the same as that for the gyrational motion of the particle; $l = -1$ corresponds to the anomalous Doppler resonance, and is appropriate when the handedness of the field and particle are opposite. Note that the transverse electric field can be decomposed into right- and left-hand circularly polarized components, so that both $l = \pm 1$ should be considered for a given wave and particle.

The frequency $\omega$ and wavevector $\vec{k}$ are related through the warm-plasma dispersion relation (e.g., Stix 1992, Chap. 10). The cold-plasma approximation, which involves neglecting the particle temperatures, results in a much simpler dispersion relation (e.g., Swanson 1989), which is also quite accurate as long as the waves are not near a cyclotron frequency (e.g., Miller & Steinacker 1992) or another natural frequency of the system, such as the electron plasma frequency $\omega_p$. In this study, we use the cold-plasma approximation and consider only parallel-propagating waves in a fully-ionized hydrogen plasma. The generation of electron plasma waves, with a parallel electric field, remains problematic (Melrose 1980, Vol. 2, pp. 52–57), and we thus treat only the circularly-polarized waves.

A discussion of these modes appears in Steinacker & Miller (1992a), but we quickly summarize them here. For $\omega \ll \Omega_p$ (where $\Omega_p$ is the proton cyclotron frequency), the shear Alfvén wave is left-hand polarized while the fast mode wave is right-hand polarized. Both have dispersion relation $\omega = v_A |k_\parallel|$. As $\omega \to \Omega_p$, $k_\parallel$ on the shear Alfvén branch approaches infinity, and waves in this regime are called ion-cyclotron waves. The fast-mode waves pass through $\Omega_p$ and become whistlers in the regime $\Omega_p \ll \omega \ll \Omega_e$, where $\Omega_e$ is the electron cyclotron frequency. The whistler dispersion relation is $k_\parallel^2 = \omega_p^2 \omega/(c^2 \Omega_e)$. As $\omega \to \Omega_e$, whistlers become electron cyclotron waves.

For parallel waves, higher-order gyroresonances cannot be satisfied and only $l = \pm 1$ are relevant. Inserting the low-frequency dispersion relation into the resonance condition (8), we see that protons can resonate with shear Alvén waves via $l = +1$, or with fast mode waves via $l = -1$, if $p \gg \beta_A/|\mu|$, where $\mu$ is the pitch-angle cosine. However, we emphasize that sub-Alfvénic protons can also resonate with waves, but of frequency above the region of applicability of the Alfvén dispersion relation. Specifically, from equation (8) it is clear that very low-energy particles will resonate with waves of frequency comparable to the particle cyclotron frequency, which, in the case of protons, are ion-cyclotron waves. Similarly, electrons can gyroresonate with fast-mode waves via $l = +1$ or shear Alfvén waves via $l = -1$ if $p \gg m_p \beta_A/m_e|\mu|$ (see also Melrose 1974; Miller & Ramaty 1987).

The effect of a spectrum of waves upon the particle phase-space distribution function $f(\vec{p})$ can be determined by solving a diffusion equation for $f(\vec{p})$ in momentum $\vec{p}(\equiv \gamma\vec{\beta})$ space. The diffusion coefficients in this equation can be readily calculated using the Hamiltonian approach of Karimabadi et al. (1992; see also Miller & Roberts 1995). However, if the pitch angle appreciably changes on a time scale much less than the acceleration time scale, then the distribution is isotropic over the latter time scale and this two-dimensional diffusion equation can be readily averaged over $\mu$ to obtain a diffusion equation in $p$-space only. In this case, acceleration is characterized by a single momentum diffusion coefficient $D(p)$, with units of $t^{-1}$. The convective and diffusive nature of the acceleration is best revealed by writing the resulting momentum diffusion equation as a Fokker-Planck equation in energy space (Tsytovich 1966; Melrose 1980, Vol. 2, p. 53–5). The convection coefficient (or systematic energy-gain rate) in the Fokker-Planck equation is related to $D(p)$ by

$$\frac{1}{mc^2}\langle\frac{dE}{dt}\rangle = \frac{1}{p^2}\frac{\partial}{\partial p}[\beta p^2 D(p)] \,, \tag{9}$$

where $E$ is the particle energy.

The proton distribution in the presence of Alvén waves can be taken to be isotropic (e.g.,

Barbosa 1979). The wave spectrum in an AGN is clearly unknown, but can be constrained by general arguments. In the cascading scenario, waves at a given wavelength will nonlinearly cascade (e.g., Achterberg 1979; Eichler 1979) to shorter wavelengths, and an injection of turbulence at long wavelengths will lead to a continuous wave spectrum extending from this injection wavelength, through the inertial range, and into the short wavelength dissipation range, where wave energy is ultimately dissipated through resonance on the background particles (see Miller & Roberts 1995 for a quantitative treatment of this). The waves are usually described by a spectral density $W(\vec{k})$, where $W(\vec{k})d\vec{k}$ is the energy density of waves with wavevector in the element $d\vec{k}$ about $\vec{k}$. For isotropic wave distributions, the cascade of spectral energy can be analytically described with either a Kolmogorov or Kraichnan phenomenology (see Zhou & Matthaeus 1990), yielding an inertial range spectral density $W(k) \propto k^{-q}$, where $k = |\vec{k}|$, $q = 3/2$ for the Kraichnan (1965) spectrum, and $q = 5/3$ for the Kolmogorov spectrum. One-dimensional MHD simulations (Miller & Roberts 1995) have shown that the Kolmogorov phenomenology is more appropriate, even for low-amplitude waves in a magnetized plasma (see Verma 1994 for a similar conclusion for two-dimensional turbulence), but we will consider both cases.

The formation of the turbulent wave spectrum could also result from wave generation in fully developed fluid turbulence through the Lighthill mechanism (Henriksen, Bridle, & Chan 1982; Eilek & Henriksen 1984). In this process, wave injection takes place over a range of $k$ which depends on the Reynolds number of the fluid, and feedback between the particle and wave energy could drive the distribution to a self-similar power-law form. In the case of Lighthill-generated waves, the spectral indices of the MHD spectrum are modified over the classical Kolmogorov & Kraichnan values of 5/3 and 3/2, respectively. The results of the present study do not depend on the origin of the turbulent wave spectrum, although we do assume that it does have a power-law form.

We first consider particle acceleration by low-frequency ($\omega \ll \Omega_p$) waves. We normalize the spectral density of low-frequency parallel waves, assuming symmetry and equipartition (i.e., $W(k_{\parallel}) = W(-k_{\parallel})$), such that

$$\zeta_i \equiv \frac{W_i^{tot}}{U_B} = \frac{2 \int_{k_{min}}^{\infty} dk_{\parallel} \, W(k_{\parallel})}{U_B} \, , \tag{10}$$

where $W_i^{tot}$ is the total energy density in mode $i$ (either left- or right-hand circularly polarized waves), which includes magnetic and electric fields as well as the bulk particle oscillation kinetic energy in the plasma. The term $k_{min}$ is the minimum wave number, which we take to correspond approximately to the inverse size scale of the system (that is, we assume that waves are produced at scales approximately equal to the size of the accretion disk). This is a very conservative assumption, and much higher acceleration rates (see below) could be obtained if we assumed that waves were generated and started cascading on smaller scales.

Using the above-mentioned Hamiltonian formalism (Miller & Roberts 1995; see also Appendix

A for a simplified derivation of the stochastic diffusion coefficients), we find that the pitch-angle averaged momentum-diffusion coefficient for ions resonating with shear Alfvén waves, or electrons resonating with fast mode waves, is given by

$$D(p) = \frac{\pi}{2} \left[ \frac{q-1}{q(q+2)} \right] (ck_{min})\beta_A^2 \zeta_i (r_L k_{min})^{q-2} \frac{p^q}{\beta} \, , \tag{11}$$

where $r_L \equiv mc^2/eB$ is the nonrelativistic Larmor radius of the particle. The quantity $\zeta_i$ is the normalized energy density in either shear Alfvén or fast mode waves. This result can also be derived from the results given by Melrose (1974), Miller & Ramaty (1989), or Bogdan, Lee, & Schneider (1991). Using equation (9), the systematic energy-gain rate is

$$\langle \frac{d\gamma}{dt} \rangle = \frac{\pi}{2} \left( \frac{q-1}{q} \right) (ck_{min})\beta_A^2 \zeta_i (r_L k_{min})^{q-2} p^{q-1} \, , \tag{12}$$

so that the characteristic time scale for reaching Lorentz factor $\gamma$, letting $k_{min} \to R^{-1}$, is given by

$$t_E \equiv \left| \frac{1}{(\gamma-1)} \langle \frac{d\gamma}{dt} \rangle \right|^{-1} = \frac{2}{\pi} \left( \frac{q}{q-1} \right) \frac{t_{dyn}}{\beta_A^2 \zeta_i} \left( \frac{r_L}{R} \right)^{2-q} \left( 1 - \gamma^{-1} \right) \frac{p^{2-q}}{\beta} \, . \tag{13}$$

Since the low-frequency dispersion relation was employed in the derivation of the diffusion coefficient leading to equations (12) and (13), they are consequently valid only for ions and electrons above the minimum momenta discussed earlier. A correct treatment of ion acceleration at lower energies, where resonance is with waves of frequency above the region of applicability of the Alfvén dispersion relation, is very complicated and the coefficient must be numerically computed (Steinacker & Miller 1992b). However, equation (11) remains a fair approximation even at low (sub-Alfvénic) proton energies. While electrons with energies below the threshold specified above no longer resonate with Alfvén waves, they can resonate with higher frequency waves, namely whistlers. We consider these waves in § 2.3. For high energy particles, the absolute maximum energy is limited by the size of the system, so that we require $r_L < R$. Using the results from § 2, we find that $\gamma \lesssim 5 \times 10^{11} (m_p/m) B_3 R_2$, where $B_3$ is the magnetic field in $10^3$ G. We show below, however, that diffusive escape and competition with other radiation processes typically limit the particle energy to much lower values than this.

Pitch-angle scattering controls the diffusion of particles in physical space, and therefore the diffusive escape time, $T_d$, from the system. Using the same procedure mentioned above, we find the pitch-angle scattering diffusion coefficient for Alfvén turbulence to be

$$D_{\mu\mu} = \frac{\pi(q-1)}{4} (ck_{min})\zeta_i (r_L k_{min})^{q-2} \beta p^{q-2} |\mu|^{q-1} (1 - \mu^2) \, , \tag{14}$$

(see also Miller & Ramaty 1989). This coefficient determines the rate of diffusion along the magnetic field through the expression

$$\kappa_\parallel^i = \frac{\beta^2 c^2}{8} \int_{-1}^{1} d\mu \, \frac{(1-\mu^2)^2}{D_{\mu\mu}^i} \,, \tag{15a}$$

(Earl 1974; Schlickeiser 1989), where $\kappa_\parallel$ is the parallel diffusion coefficient. Assuming that the magnetic field lines are directed radially outward, the escape time is given by

$$T_d \approx \frac{R^2}{4\kappa_\parallel} \tag{15b}$$

(Barbosa 1979; Steinacker & Miller 1992a), and thus we obtain

$$T_d \approx \frac{\pi}{8} \, (q-1)(2-q)(4-q) \, t_{dyn} \, \zeta_i \, (r_L k_{min})^{q-2} \, \frac{p^{q-2}}{\beta} \,, \tag{16}$$

which is valid for $1 < q < 2$. Equation (16) is based on normalization (10) for the wave energy density, where we assume that the turbulent wave spectrum is described by a single power law with maximum wave vector $k_{max} \to \infty$. Rederiving equation (16) for arbitrary values of $k_{max}$ is straightforward, and generalizes the expression for values of $q > 2$. The time scale for diffusive escape of electrons must also include whistler interactions, however, and is treated in § 2.3.

To get acceleration to momentum $p$, it is necessary that the diffusive escape time scale exceed the dynamical time scale. Otherwise, the particles will escape directly from the system without significant acceleration. In fact, equation (16) is unphysical if $T_d < t_{dyn}$. The condition $T_d > t_{dyn}$ implies that only particles with

$$\beta^{\frac{1}{2-q}} p \lesssim \left(\frac{R}{r_L}\right) \left[\frac{\pi}{8}(q-1)(2-q)(4-q)\zeta_i\right]^{\frac{1}{2-q}} \tag{17}$$

can be accelerated prior to leaving the system. From the definitions for $R$ and $r_L$, and the expression for $B$ from equation (4), we have

$$\left(\frac{R}{r_L}\right) = 2.92 \times 10^{11} \left(\frac{m_p}{m}\right) b M_8^{1/2} l_{Edd}^{1/2} \,, \tag{18}$$

where $m = m_p$ for proton acceleration and $m = m_e$ for electron acceleration. Using equations (17) and (18), we find that for the Kraichnan spectrum, $q = 3/2$,

$$\gamma \lesssim 1.8 \times 10^8 \left(\frac{m_p}{m}\right) \zeta_{-1}^2 b M_8^{1/2} \ell_{Edd}^{1/2} \,, \tag{19a}$$

where $\zeta_i = 0.1\zeta_{-1}$. For the Kolmogorov spectrum, $q = 5/3$, we find that

$$\gamma \lesssim 2.5 \times 10^6 \left(\frac{m_p}{m}\right) \zeta_{-1}^3 b M_8^{1/2} \ell_{Edd}^{1/2} . \tag{19b}$$

There is no solution to equation (17) when $q = 2$.

Acceleration to momentum $p$ requires, in addition, that $t_E < T_d$, so that particles do not escape before being accelerated to this momentum. From equations (13) and (16), we find that this implies

$$p^{2-q} \lesssim \frac{\pi}{4} (q - 1) \left[\frac{(2-q)(4-q)}{q}\right]^{1/2} \left(\frac{R}{r_L}\right)^{2-q} \beta_A \zeta_A . \tag{20}$$

For the case $q = 3/2$, we find that

$$\gamma \lesssim 7.4 \times 10^6 \left(\frac{m_p}{m}\right) \frac{\zeta_{-1}^2 b^3 M_8^{1/2} l_{Edd}^{3/2}}{\tau_p R_2} \tag{21a}$$

using equation (5) for $\beta_A$. When $q = 5/3$, we find that

$$\gamma \lesssim 3.7 \times 10^4 \left(\frac{m_p}{m}\right) \frac{\zeta_{-1}^3 b^4 M_8^{1/2} l_{Edd}^2}{(\tau_p R_2)^{3/2}} . \tag{21b}$$

There is no solution to equation (20) when $q = 2$. Neglecting radiative losses, we therefore find that the most stringent constraint on the maximum particle energy is given by a comparison of acceleration and diffusive escape time scales for the typical parameters we encounter in this study.

### 2.3. Stochastic Acceleration of Electrons by Whistler Turbulence

We next consider the acceleration of particles due to whistler turbulence. Since the whistler electric field rotates in the right-handed sense while the proton gyrates in the left-handed sense, the relevant harmonic number is $l = -1$. From the whistler dispersion relation and the resonance condition, we see that protons can resonate with whistlers only for a narrow range of energies and pitch angles. Consequently, we can neglect proton acceleration by whistlers.

Electrons, on the other hand, rotate in the same sense as the whistler electric field, and thus interact through $l = +1$. Using the whistler dispersion relation as well as the resonance condition, we see that electrons can resonate provided that $\beta_A (m_p/m_e)^{1/2} < p < \beta_A (m_p/m_e)$ (see Melrose 1974). However, lower energy electrons can still be accelerated through resonant interactions by waves with frequencies higher than whistlers, but the diffusion coefficient and acceleration rate in this regime are now more complicated (Steinacker & Miller 1992a). Similarly, acceleration of higher energy electrons is also possible, but is now accomplished by fast mode waves, as described in the previous section.

Again using the Hamiltonian technique, we find after much algebra that the pitch-angle averaged momentum diffusion coefficient for electrons and right-handed waves (which include waves from the fast mode through the whistler regime) is given by

$$D(p) = \frac{\pi}{4} (q-1)(\frac{m_p}{m_e})^2 (ck_{min}) \frac{\beta_A^4}{\beta} \zeta_R (r_L k_{min})^{q-2} p^{q-2} \left[ \frac{(p\kappa_1)^{2-q}}{2-q} + \frac{(p\kappa_1)^{-q}}{q} - \frac{2}{q(q-2)} \right] \;, \quad (22)$$

where $\kappa_1 = (43\beta_A)^{-1}$ is the maximum wavenumber in the whistler spectrum, obtained by assuming that the whistler dispersion relation extends up to $\Omega_e$. Due to the more complicated properties of whistlers, the normalized turbulence energy density $\zeta_R$ only contains the contribution from the wave magnetic field; including the contribution from the electric field as well as from the bulk kinetic energy of the background particles will result in a diffusion coefficient lower by a factor of at most a few. The systematic energy-gain rate is obtained from equation (22) using equation (9), giving

$$\langle \frac{d\gamma}{dt} \rangle = \frac{\pi}{2} \frac{(q-1)}{(2-q)} (\frac{m_p}{m_e})^2 (ck_{min}) \beta_A^4 \zeta_R (r_L k_{min})^{q-2} p^{q-3} [(p\kappa_1)^{2-q} - 1] \;, \quad (23)$$

so that the time scale for reaching Lorentz factor $\gamma$, again letting $k_{min} \to R^{-1}$, is given by

$$t_E = \frac{2}{\pi} (\frac{m_e}{m_p})^{1+q/2} (\frac{q-2}{q-1}) \frac{t_{dyn}}{\beta_A^{2+q} \zeta_R} (\frac{r_L}{R})^{2-q} \frac{p^2}{\beta} (1 - \gamma^{-1}) [(43\beta_A/p)^{2-q} - 1]^{-1} \;. \quad (24)$$

The acceleration time scales for electrons interactions with whistlers, equation (24), and low-frequency fast-mode waves, equation (13) are not continuous, due to the discontinuity in the dispersion relations. Employing the correct dispersion relation for right-handed waves from low to high frequencies (Steinacker & Miller 1992a) is necessary in order to obtain results throughout the whistler and fast-mode regime. The derivation of this result is presented in Appendix B.

The pitch-angle scattering diffusion coefficient of electrons interacting with whistlers is given by

$$D_{\mu\mu} = \frac{\pi}{4}(q-1)(ck_{min})\zeta_R (r_L k_{min})^{q-2} \beta p^{q-2} |\mu|^{q-1} (1 - \mu^2) \;. \quad (25)$$

Following the procedure for the Alfvén waves, we find that the diffusive escape time scale is given by

$$T_d \cong \frac{\pi}{4}\,(q-1)\,t_{dyn}\,\zeta_R\,(r_L k_{min})^{q-2}\,\frac{p^{q-2}}{\beta} \times \left[\frac{(p/43\beta_A)^{q-2}-1}{q-2} - \frac{(p/43\beta_A)^{q-4}-1}{q-4}\right]^{-1}. \qquad (26)$$

Equation (26) generalizes equation (16) for electrons to include both the Alfvén and whistler interactions, and reduces to equation (16) in the limit $p \gg \beta_A(m_p/m_e)^{1/2}$.

## 3.  Particle Acceleration and Energy Losses in Black Hole Magnetospheres

### 3.1.  Proton Acceleration and Energy Losses

Besides escape from the system, particle acceleration will also be limited by energy losses. The most important energy-loss processes for high-energy protons are secondary pion production from collisions with thermal protons in the corona, and through photo-pair and photo-pion production when energetic protons interact with photons radiated by the accretion disk. We estimate the energy loss rate of protons through proton-proton collisions by the expression

$$-(\frac{d\gamma}{dt})_{pp} \cong \eta\sigma_{pp}n_p c\gamma \qquad (27)$$

where, in the high-energy limit $\gamma_p \gg 1$, the inelastic proton-proton cross section $\sigma_{pp} \approx 30$ mb and the inelasticity $\eta \approx 1/2$ (e.g., Gaisser 1990). Using equations (2) and (6), we obtain

$$\frac{t_{pp}}{t_{dyn}} \cong \frac{44}{\tau_p} \qquad (28)$$

for the ratio of the energy-loss time scale $t_{pp}(\equiv 1/n_p\sigma_{pp}\beta c)$ though p-p collisions to the dynamical time scale. Because the threshold for pion production is $\approx 300$ MeV, $t_{pp}^{-1} \to 0$ when $p \lesssim 0.9$.

The tabulated values of Begelman, Rudak, & Sikora (1990) are used to determine the energy-loss time scales for photo-pair and photo-pion production. We approximate the disk radiation field by a blackbody spectrum multiplied by a graybody factor representing the ratio of the photon energy density given by equation (3) to the blackbody photon energy density at temperature $T = m_e c^2\theta/k_B$, where $k_B$ is Boltzmann's constant. For the effective temperature of the disk, we consider two values which should encompass the range of likely radiation-field temperatures. If the "blue bump" emission observed from a wide variety of AGNs is interpreted as the thermal emission from an optically thick, geometrically thin accretion disk, then temperatures $k_B T \sim 50$ eV are likely. Hard X-ray and soft gamma radiation with color temperatures $k_B T \sim 50$ keV is also seen in the spectra of Seyfert AGNs (Maisack et al. 1993; Johnson et al. 1994), and is probably produced by a hot, optically thin plasma in the inner regions near the central black holes. Thus we let $\theta = 10^{-4}$ or $10^{-1}$ in our calculations. The radiation field in the central regions of AGNs would probably be more accurately described by a superposition of graybodies with temperatures in the range $\sim 50$ eV - 50 keV.

For the energy-loss rate of protons through Coulomb interactions, we use equation (4.22) of Mannheim & Schlickeiser (1994), written as a Coulomb energy-loss time scale $t_{Coul}$ through the expression

$$\frac{t_{Coul}}{t_{dyn}} \cong 49 \, \frac{\left(3.8\theta_{pl}^{3/2} + \beta^3\right)(\gamma - 1)}{\tau_p \, \beta^2 \Lambda_{25}} \, , \tag{29}$$

where $m_e c^2 \theta_{pl}/k_b$ is the temperature of the thermal plasma in the corona and the Coulomb logarithm $\Lambda$ is given by $\Lambda = 25\Lambda_{25}$. The ratio of the proton synchrotron loss time scale to the dynamical time scale is given by

$$\frac{t_{p,syn}}{t_{dyn}} \cong \frac{2.5 \times 10^8}{\gamma} \, \frac{R_2}{b^2 \ell_{Edd}} \, , \tag{30}$$

and is always smaller than other loss processes for the parameters considered. We therefore do not consider it further.

In Figure 2, we plot the various time scales for acceleration, escape, and energy loss as a function of dimensionless proton momentum $p_p = \beta_p \gamma_p$. Here we choose a black hole mass $M = 10^8 M_\odot$, a coronal radius $R = 10^2 R_g$, and take $\tau_p = 1$. We also assume that the thermal coronal plasma temperature is equal to $0.1 m_e c^2$. The energy loss rates for photo-pair and photo-pion production are given for both $\theta = 10^{-4}$ and $\theta = 10^{-1}$. The diffusive escape and acceleration time scales given by equations (13) and (16) are only estimates below $p_p \cong \beta_A$, because the waves that are resonant with particles at these momenta have frequencies in the range where the Alfvén dispersion relation is no longer valid. We therefore limit our calculations to proton momenta greater than this value.

In Figure 2a, we show results where the ratio of shear Alfvén wave energy density to magnetic field energy density, $\zeta_A$, is 10%, roughly corresponding to the upper limit where quasilinear theory applies (e.g., Zachary 1987). Here we assume a Kolmogorov spectrum for the turbulence, so that $q = 5/3$. We also assume an equipartition magnetic field ($b = 1$) and let the system be accreting at the Eddington limit ($l_{Edd} = 1$). As can be seen, protons are accelerated to $\gamma_p \approx 4 \times 10^4$ and then diffusively escape from the system. Only a small fraction of energy is radiated prior to escape, primarily in the form of photo-pair production if the radiation temperature is near 50 keV, or in the form of secondary pion production from interactions with the thermal background. Most of the accelerated particle energy is lost through bulk kinetic energy in the outflowing particles. If there is efficient damping of the waves through resonant interactions with the accelerated protons, this means that nearly 10% of the accretion energy is radiated in the form of accelerated particles escaping from the system.

Figure 2b shows the result of using a Kraichnan spectrum for the wave turbulence, with all other parameters the same as in Figure 2a. Because the total turbulent energy density is the same in the two cases, there is more turbulent energy per unit wave number at large values of $k$ for the flatter Kraichnan spectrum. Because plasma waves with larger wave numbers preferentially resonant with lower energy particles, the time scale for acceleration of lower-to-moderate energy protons is therefore more rapid for the Kraichnan spectrum than the Kolmogorov spectrum. Instead of flowing out of the system, we see in this case that the increased turbulence impedes diffusive escape and increases the escape time scale sufficiently that particle acceleration is halted by energy losses through photo-pion production.

Figure 2c shows the results of lowering the Eddington luminosity by two orders of magnitude, with all other parameters the same as in Figure 2a. The maximum energy to which protons can be accelerated before escaping from the system is reduced by $\sim 4$ orders of magnitude in Fig. 2c compared with Fig. 2a, in agreement with equation (21b). Acceleration of nonthermal protons is nearly halted at $p_p = 0.6$ or at proton kinetic energies $\approx 150$ MeV because the Coulomb energy-loss time scale is almost equal to the acceleration time scale. In Figure 2d, stochastic acceleration of protons to high energies is clearly impossible due to dominant Coulomb losses. All parameters are the same as in Figure 2a in this plot, except that the energy density of plasma turbulence is assumed to be 1% of the energy density in the large scale magnetic field ($\zeta_A = 1\%$), and the strength of the large scale field is taken to be only 10% of its equipartition value ($b = 0.1$). Virtually all wave energy goes into heating the background plasma, and no energy is dissipated through bulk particle outflow. In Figure 2e, we consider a $10^9$ M$_\odot$ black hole accreting well below the Eddington limit, with the dissipation taking place in a very dilute corona with $\tau_p = 0.01$. Because of the low density corona, the Coulomb barrier does not prevent particle outflow. Such a system may be relevant to observations of M87, as we discuss in more detail below.

The parameter values at which the Coulomb barrier prevents proton acceleration to high energies can be obtained by setting the Coulomb energy-loss time scale (29) equal to the acceleration time scale (13). For the case $q = 5/3$, one finds that the maximum value of the term $\beta^2 p^{-2/3}$ as a function of $p$ corresponds closely to the proton momentum where the acceleration and Coulomb loss rates are equal, provided that the term $3.8\theta_{pl}^{3/2}$ is small in comparison with $\beta_o^3$. This maximum occurs at $p_o = 2^{1/2}$ or $\beta_o = (2/3)^{1/2}$. If $\theta_{pl} \ll 0.3$, one therefore obtains the following relation among parameter values defining the regimes where proton acceleration to relativistic energies is not possible, given by

$$\Lambda_{25}\tau_p^2 R_2 M_8^{-1/6} b^{-7/3} \ell_{Edd}^{-7/6} \zeta_{-1}^{-1} \gtrsim 500 \, . \tag{31a}$$

Equation (31a) holds for $q = 5/3$; an analogous derivation gives

$$\Lambda_{25}\tau_p^2 R_2 M_8^{-1/4} b^{-5/2} \ell_{Edd}^{-5/4} \zeta_{-1}^{-1} \gtrsim 3.1 \times 10^4 \qquad (31b)$$

for $q = 3/2$. Equation (31a) show that as $l_{Edd}$ becomes greater than $\approx 0.005$, with all other parameters taking their standard values, there is a transition between a system in which there is no nonthermal proton acceleration to one in which there is a significant luminosity in nonthermal protons. This value is in general agreement with the graphical results which suggest that the transition is at $l_{Edd} \cong 0.008$ (see Fig. 2c). Alternately, one can vary $b$ or $\zeta_{-1}$ to find that as the magnetic field is decreased from its equipartition value, or the wave energy density is decreased from the fully turbulent regime, the system undergoes the same sort of transition. For other parameters assigned their nominal values, we find from equation (31a) that the transition occurs at $b \approx 0.07$ or $\zeta_{-1} = 0.002$. We consider the significance of these results in § 4.

### 3.2. Electron Acceleration and Energy Losses

The acceleration of electrons to high energies in an accretion-disk corona will be limited principally by Coulomb, bremsstrahlung, Compton and synchrotron losses. Higher-order processes, such as triplet pair production or radiative ("double") Compton scattering, are generally much less important in the environments of AGNs, and will not be considered here. For the Coulomb energy-loss time scale, we use the simple expression $-\dot{\gamma}_{Coul} = 4\pi n_p r_e^2 c \ln \Lambda / \beta$ (see Gould 1972 for accurate expressions for the Coulomb logarithm). In terms of the dynamical time scale, we therefore have

$$\frac{t_{Coul}}{t_{dyn}} \cong \frac{2(p - \beta)}{75\tau_p \Lambda_{25}} , \qquad (32)$$

which is valid for $\beta \gg (3\theta_e)^{1/2}$. Bremsstrahlung losses dominate Coulomb losses only when $\gamma \gg 1$. Using the relativistic bremsstrahlung energy loss rate for a fully ionized hydrogen plasma (e.g., Blumenthal & Gould 1970), we obtain

$$\frac{t_{ff}}{t_{dyn}} \cong \frac{\pi}{3\alpha_f \tau_p [\ln(2\gamma) - 1/3]} , \qquad (33)$$

where $\alpha_f$ is the fine-structure constant.

The Thomson energy-loss time scale $t_T$ given by equation (7) is only valid when $\gamma \ll \theta^{-1}$, where $\theta$ is the color temperature of the radiation field. The ratio of $t_T$ to the dynamical time scale $t_{dyn}$ can be rewritten as

$$\frac{t_T}{t_{dyn}} \cong \frac{0.04 R_2}{(1 + \gamma) l_{Edd}} . \qquad (34)$$

We use equation (2.59) of Blumenthal & Gould (1970) for the energy-loss rate of an electron scattering thermal photons in the extreme Klein-Nishina limit, corrected by a graybody factor

$$\chi = \frac{l_{Edd}L_{Edd}}{4\pi R^2 \sigma_{SB} T^4} \simeq \frac{6.5 \times 10^{-21} l_{Edd}}{R_2^2 M_8 \theta^4} \,, \tag{35}$$

where $\sigma_{SB}$ is the Stefan-Boltzmann constant. We obtain

$$\frac{t_{KN}}{t_{dyn}} \simeq \frac{6 r_e (\gamma - 1)}{\chi \pi \alpha_f^3 R \theta^2 \ln(0.552\gamma\theta)} \simeq \frac{0.14 R_2 \theta^2 (\gamma - 1)}{l_{Edd} \ln(0.552\gamma\theta)}, \tag{36}$$

valid for $\gamma \gg \theta^{-1}$. Equations (34) and (36) are smoothly connected in the intermediate regime $\gamma \approx \theta^{-1}$ using a simple bridging formula.

The synchrotron energy-loss time scale is easily derived from the pitch-angle averaged energy-loss rate, giving

$$\frac{t_{syn}}{t_{dyn}} \simeq \frac{0.04 R_2}{(\gamma + 1) b^2 l_{Edd}} \,. \tag{37}$$

Equation (37) is a lower limit to the synchrotron-loss time scale, because nonthermal synchrotron self-absorption reduces the electron energy-loss rate. For a power-law electron spectrum given by $n(\gamma) = K\gamma^{-p}$, where $n(\gamma)d\gamma$ is the number density of electrons with $\gamma$ between $\gamma$ and $\gamma + \mathrm{d}\gamma$, the synchrotron self-absorption coefficient is given by

$$\kappa[\mathrm{cm}^{-1}] = \frac{\sigma_T K}{16 \cdot 3^{1/2} \alpha_f \epsilon_B} \Gamma(\frac{3p+2}{12})\Gamma(\frac{3p+22}{12})(\frac{3\epsilon_B}{\epsilon})^{\frac{p+4}{2}} \tag{38}$$

(Rybicki & Lightman 1979). Here $\epsilon_B$ is the magnetic field in units of the critical field $B_{cr} = m_e^2 c^3/e\hbar = 4.414 \times 10^{13}$ G, and we have assumed a $90°$ pitch angle for the electrons. Solving for the self-absorption energy $\epsilon_{SSA}$ which solves $\kappa R = 1$, and noting that self-absorption effects change the energy-loss rate for electrons with $\gamma \lesssim \bar{\gamma} \cong (\epsilon_{SSA}/\epsilon_B)^{1/2}$, we can derive an approximate upper limit for $\bar{\gamma}$ by assuming that the normalization $K$ is determined by the condition that all wave power is dissipated as synchrotron radiation. The result is

$$\bar{\gamma} \approx 3^{1/2} \left[\frac{8 \times 10^{11}(3-p)\zeta R_2 M_8^{1/2}}{b^3 l_{Edd}^{1/2} \gamma_{max}^{3-p}} \Gamma(\frac{3p+2}{12})\Gamma(\frac{3p+22}{12})\right]^{\frac{1}{p+4}} \,, \tag{39}$$

where $\gamma_{max} \sim 10^2\text{-}10^3$ is the maximum Lorentz factor in the electron spectrum, which should be determined self-consistently. The value $p = f(q) \approx 1$ for stochastic acceleration of electrons by whistler turbulence (Li 1995). By substituting in standard values in equation (39), we find that self-absorption can severely reduce the synchrotron energy loss rate for electrons with $\gamma \ll 10^2\text{-}10^3$.

Thus the synchrotron loss time scale for low-energy electrons can be much longer than given by equation (37).

Keeping in mind that the synchrotron loss time scale (37) is a lower limit, we plot in Figure 3 the acceleration, diffusive escape, and energy loss-time scales for electrons interacting with whistler and fast mode waves in an accretion-disk corona. We truncate the acceleration and diffusive escape time scales at $\rho \lesssim 2\beta_A(m_p/m_e)^{1/2}$ because lower-momenta electrons with momenta in this range are in resonance with waves outside the the whistler regime. Such electrons may still resonate with waves with higher wave numbers, namely electron cyclotron waves, but the diffusion coefficients are different than those presented in §2.3. The time scales probably smoothly extend to the lower momenta particles found in the tail of the thermal distribution, but finite temperature effects make an analytic derivation untenable (see Steinacker & Miller 1992a).

Figures 3a and 3b use the same parameters as in Figures 2a and 2b for protons. The acceleration time scale in both cases is much shorter than the Coulomb/bremsstrahlung ("ff") time scale, so electrons will be accelerated out of the background plasma if we can assume that the time scales extend smoothly to lower momenta. Compton or synchrotron losses limit the electron acceleration to higher energies, depending on the mean photon energy. Because Klein-Nishina effects reduce the energy loss rate, lower-temperature radiation fields are more effective at halting electron acceleration for a given energy density. Depending on the magnetic field strength, synchrotron losses may dominate Compton losses, although a self-consistent treatment of synchrotron self-absorption is required to accurately specify the maximum electron energy where the synchrotron energy-loss rate balances the electron energy-gain rate.

Figure 3c shows the results of reducing the accretion luminosity by two orders of magnitude, with all other parameters the same as in Figure 3a. Because the acceleration time scale increases with decreasing accretion luminosity, the Coulomb/ff time scale at low electron momenta is nearly equal to the acceleration time scale. At sufficiently low values of $l_{Edd}$, acceleration is halted. We can derive an relation among parameter values which approximately determines whether electrons are accelerated over the Coulomb barrier, using equation (32) and assuming that the acceleration time scale (24) can be extended to low momenta (by replacing the term in brackets in eq.[24] by unity). For $q = 5/3$, electron acceleration is halted when

$$\Lambda_{25}\tau_p^{7/3}R_2^{4/3}M_8^{-1/6}b^{-3}\ell_{Edd}^{-3/2}\zeta_{-1}^{-1} \gtrsim 10^4 \ . \tag{40a}$$

For $q = 3/2$, we find that the Coulomb barrier halts acceleration of electrons to high energies when

$$\Lambda_{25}\tau_p^{9/4}R_2^{5/4}M_8^{-1/4}b^{-3}\ell_{Edd}^{-3/2}\zeta_{-1}^{-1} \gtrsim 10^6 \ . \tag{40b}$$

## 4. Discussion

Galactic black hole candidates and active galactic nuclei display a wide range of phenomena, including luminous variable X-ray emission, UV bumps and soft X-ray excesses, radio and gamma-ray jets, intraday variability, superluminal motion, etc. Yet in all cases the ultimate energy source powering the emission is thought to be the conversion of the gravitational potential energy of accreting matter into radiation. The radiation is dissipated in the process of transporting the angular momentum of accreting matter outward. The conventional prescription for the poorly understood mechanism of angular momentum transport is through the $\alpha$ parameter, which relates the viscous shear stress to the pressure (see Pringle 1981 for a review). Magnetic viscosity through shear amplification and reconnection of entrained magnetic fields (Shakura & Sunyaev 1973; Eardley & Lightman 1975; Tagger, Pellat, & Coroniti 1992) provides one such mechanism. Although we do not treat the underlying microphysics of the global angular momentum transport here, generation of magnetic turbulence in an accretion-disk corona would accompany the dissipation of angular momentum through magnetic viscosity.

We have shown that the existence of an accretion-disk corona with a large-scale ordered magnetic field supporting a spectrum of plasma waves can accelerate particles to high energies. Calculations of the detailed radiation signatures of this system for various parameter values are deferred to a later paper, but we can qualitatively describe the possible spectral states implied by stochastic particle acceleration. Our principal result is that the Coulomb barrier separates systems which display nonthermal particle acceleration and outflow from those which do not.

### 4.1. No Acceleration over the Coulomb Barrier

If the Coulomb barrier prevents acceleration of protons to relativistic energies, then the wave energy is dissipated as heat. Because the Compton and synchrotron energy-loss rates of leptons exceeds that of ions, a two-temperature plasma will be formed unless additional plasma processes exist to equilibrate the electron and ion temperatures. Compton cooling of electrons by a soft photon source produces a Sunyaev-Titarchuk (1980) spectrum at nonrelativistic electron temperatures and large optical depths; modifications at mildly relativistic temperatures and low optical depths have recently been derived (Hua & Titarchuk 1995) and calculated for plasmas in pair balance (Skibo et al. 1995). The hard X-ray spectrum of NGC 4151, a Seyfert 1.5-2 AGN, is well fit by such a spectrum (Titarchuk & Mastichiadis 1994), as is the Seyfert 1 galaxy IC 4329A, provided that reflection from a cool optically-thick medium is included (Zdziarski et al. 1994).

Secondary pion production is usually important whenever the Coulomb barrier halts nonthermal particle acceleration, as shown in Fig. 2c. Indeed, secondary production can by itself prevent proton acceleration in certain parameter regimes. Thus the formation of a two-temperature plasma will be accompanied at some level by the production of $\pi^o$ gamma-rays and charged pion-decay electrons and positrons. The cascading of these emissions will produce a hard tail on the thermal Comptonization spectrum (Dermer 1988; Jourdain & Roques 1994), such as seen in the spectrum of Cygnus X-1 (McConnell et al. 1994).

## 4.2. Acceleration over the Coulomb Barrier without Particle Escape

When protons are accelerated over the Coulomb barrier, one of two things can occur. In the first case, when the energy-loss time scale in the accretion-disk corona is shorter than the diffusive escape time scale, nonthermal protons will preferentially lose energy rather than escape. Secondary pion or photo-pion production processes (see Fig. 2b) are the most important energy-loss processes in this case, and such systems would be prolific neutrino sources. High-energy gamma rays and secondary electrons and positrons from the pion emissions would cascade to lower energies until the acceleration time scale equaled the electron energy loss-time scale, as shown in Fig. 3. Electron acceleration by plasma waves over the Coulomb barrier will also produce a quasi-monoenergetic pileup distribution (Schlickeiser 1984) of electrons and pairs through Compton scattering and $\gamma$-$\gamma$ pair production, which would follow behavior similar to that of the cascading pion-decay emissions. The buildup of nonthermal pair energy could continue until the increased number of nonthermal particles significantly dissipates the injected wave energy, thereby limiting the acceleration of further nonthermal particles. In such circumstances, numerical simulations (Li 1995) show that the system will collapse. This differs from the conclusion of Henri & Pelletier (1991) who argued that such a system would become unstable and impulsively eject pair plasma. But they assume a constant energy injection rate per particle, which is not valid if the total power in magnetic turbulence is a fixed fraction of the accretion luminosity. Moreover, they assume a constant soft photon energy density and thus a constant cooling rate, which is also not correct.

## 4.3. Acceleration over the Coulomb Barrier with Particle Escape

In the second case, nonthermal protons reach energies where the acceleration time scale equals the diffusive escape time scale, and protons flow out of the system. Jet formation through collimated outflow, as proposed here, differs from other nonthermal production mechanisms such as hydromagnetic outflow (Begelman et al. 1984), direct electric-field acceleration (e.g., Lovelace 1976), and neutron production and escape (Contopoulos & Kazanas 1995). In this scenario, the escaping protons will be collimated by the large scale magnetic field if the nonthermal particle energy density is less than the magnetic-field energy density, which is likely because $\zeta \ll 1$ and $t_E \lesssim t_{dyn}$ over a wide range of parameters, so that a large reservoir of nonthermal particle energy cannot build up. As the energetic protons diffuse out of the system, they will carry along electrons from the thermal pool through electrostatic coupling. (It is interesting to note that no reverse current problem occurs in this situation, because we are treating particle acceleration through stochastic diffusion in momentum and physical space rather than through the application of an external electric field.)

The electrons will have the same Lorentz factor as the protons, and Compton drag of the electrons will produce a Comptonized gamma-ray spectrum which is subject to $\gamma$-$\gamma$ pair-production opacity. The produced pairs will no longer be electrostatically coupled to the protons, and can cool through Compton drag to low energies and annihilate. The emission spectrum of such a system has been considered by Coppi, Kartje, & Königl (1993), and the annihilation spectrum of the pair jet by Böttcher & Schlickeiser (1995). Doppler-boosted annihilation radiation could be the mechanism producing the peaked emission spectra in the MeV range from the "MeV blazars" GRO J0516-609 (Bloemen et al. 1995) and PKS 0208-512 (Blom et al. 1995) discovered with Comptel. Because of incomplete Compton cooling of the outflowing proton/electron plasma, energy will be transported to large distances from the central nucleus, and could produce the large-scale radio jets and lobes seen in radio galaxies after reconversion of the bulk kinetic energy to nonthermal particle energy through interactions with the surrounding medium.

The triggering of nonthermal particle outflow is sensitive to the values of $\zeta$, $b$, $\tau_p$, and $l_{Edd}$. If all parameters except $l_{Edd}$ are fixed, then increases in $l_{Edd}$ in a given source would be accompanied by particle outflow and the formation of radio jets. The recent discovery that X-ray transient events in the galactic black hole binary GRO J1655-40 precedes the emergence of outflowing radio-emitting plasma (Harmon et al. 1995; Hjellming & Rupen 1995) is in accord with this scenario. Similar transient behavior may accompany the superluminal galactic source GRS 1915+105 (Mirabel & Rodríguez 1994; Harmon et al. 1994). If the generation of high-energy gamma-ray flares signifies an accretion event, then the recent report of a gamma-ray flare preceding the emergence of a new radio component in 3C 279 (Wehrle et al. 1994) also indicates that accretion-rate increases trigger events of particle outflow.

More generally, we speculate that classes of objects with enhanced accretion rates will preferentially display bulk outflow. Analysis of IRAS results show that galactic interactions and mergers enhance starburst activity and the production of stellar and galactic winds (Sanders et al. 1988). These winds provide fuel that can be driven into the central cores of merging galaxies to fuel quasars. The association of elliptical galaxies with radio-loud sources is explained if elliptical galaxies are the product of mergers (Kormendy & Sanders 1992), and if fueling at Eddington-limited accretion rates is associated with particle outflow and radio activity. Thus we argue that the association of radio-loud galaxies with ellipticals and radio-quiet galaxies with spirals is a consequence of Eddington-limited accretion rates driven by galaxy mergers in the radio-loud ellipticals. The increasing fraction of radio-loud QSOs to the total number of QSOs at the bright end of the local QSO optical luminosity function (Padovani 1993) is in accord with this interpretation, although the statistics are poor. Wilson & Colbert (1995) interpret the data differently, and propose instead that inspiralling supermassive black holes accounts for the difference between radio-loud and radio-quiet AGNs. Merging of massive black holes on time scales less than the Hubble time has not, however, been demonstrated (Begelman, Blandford, & Rees 1980; Governato, Colpi, & Maraschi 1994).

The complex interplay between the various parameters in our model unfortunately prevents concrete predictions for specific objects. For example, dynamical evidence from the *Hubble Space Telescope* implies that the radio-loud galaxy M87 harbors a black hole with mass $M \cong 3 \times 10^9 M_\odot$ (Ford et al. 1994; Harms et al. 1994). This would appear to be inconsistent with the proposed scenario, insofar as the bolometric luminosity of M87 is $\approx 10^{43}$ ergs s$^{-1}$, implying that $l_{Edd} \sim 10^{-4}$. If the Thomson depth of the corona is sufficiently small, however, then particle outflow can still take place, as shown in Fig. 2e. Thus the predictive power of our scenario applies only to the statistics of a class of objects, or to the time-dependent behavior of a given object whose accretion rate is monitored through changes in its bolometric luminosity.

## 5.  Summary

We have shown that stochastic acceleration of particles by stochastic gyroresonant acceleration in accreting plasma can accelerate particles to high energies. Nonthermal particle outflow is produced when particles are accelerated over the Coulomb barrier and diffusively escape from the system. Spectral differences between classes of objects or spectral changes of the same object at different epochs are due, in this scenario, to different levels of plasma wave turbulence caused by variations of the accretion rate or changes in the strength of the magnetic field in the accretion plasma. Comparison of spectral predictions with observations will help establish the nature of the acceleration processes operating in the central engines of these sources.

C.D. thanks Reinhard Schlickeiser for spurring his interest in this subject and for discussions on various aspects this work. Comments and suggestions by Amir Levinson, Ari Laor, and the anonymous referee are gratefully acknowledged.

## A.  Intuitive Derivation of Stochastic Acceleration Diffusion Coefficients

The functional dependences of the stochastic acceleration diffusion coefficients can be derived from simple arguments. A particle's pitch angle $\phi$ changes by $\sim (\delta B/B)$ over its gyroperiod $t_g = \gamma r_L/c$, where $(\delta B/B)$ is the relative amplitude of the resonant scattering waves. At the resonant wavenumber $k$, $(\delta B/B)^2 \approx kW(k)/U_B$, and the normalization (10) of the wave energy density implies $W(k) = (q-1)\zeta_i k_{min}^{q-1} U_B k^{-q}/2$. Thus the diffusion coefficient for pitch angle scattering

$$D_{\mu\mu} \approx \frac{(\Delta\mu)^2}{t_g} \approx \left(\frac{\delta B}{B}\right)^2 \frac{c}{\gamma r_L}(1-\mu^2) \approx \frac{1}{2}(q-1)(ck_{min})\zeta_i(r_L k_{min})^{q-2}\beta p^{q-2}|\mu|^{q-1}(1-\mu^2) , \quad \text{(A1)}$$

where we substitute $k \approx (pr_L\mu)^{-1}$ using the resonance condition (8) in the limit $\omega \ll \Omega$. The factor $(1-\mu^2)$ relates the change in $\phi$ to the change in $\mu = \cos\phi$. Except for factors of order unity, equation (A1) is the same as equations (14) and equation (25).

The momentum diffusion coefficient $D_{pp}$ can be similarly derived by noting that a particle increases its momentum by $\sim \gamma \beta_g$ when it executes a complete cycle in stochastic acceleration, so that $\Delta p \sim \gamma \beta_g \Delta \mu$. Here $c\beta_g$ is the group velocity of the resonant wave. Thus

$$D_{pp} \approx \frac{(\Delta p)^2}{t_g} \approx (\gamma \beta_g)^2 D_{\mu\mu} \approx \frac{(q-1)}{2}\, \gamma^2 \beta_g^2 (ck_{min}) \zeta_i (r_L k_{min})^{q-2} \beta p^{q-2} |\mu|^{q-1} (1 - \mu^2)\,. \quad (A2)$$

For Alfvén waves, $\beta_g = \beta_A$. The pitch-angle averaged momentum diffusion coefficient $D(p) = \frac{1}{2} \int_{-1}^{+1} d\mu D_{pp}$, giving

$$D_A(p) \approx \frac{(q-1)}{2q(q+2)}\, (ck_{min}) \beta_A^2 \zeta_i (r_L k_{min})^{q-2} \frac{p^q}{\beta}\,, \quad (A3)$$

in agreement with equation (11) within a numerical factor. For Whistler waves, $\beta_g = c\Omega_e k/\omega_p^2 = m_p \beta_A^2/(m_e p \mu)$. After substituting this into equation (A2) and integrating over $\mu$, it is then straightforward to recover equation (22) within a numerical factor, noting that $|\mu| > (m_p/m_e)^{1/2} \beta_A/p$ in accordance with the assumption that the whistler dispersion relation is only valid for $\Omega_p < \omega < \Omega_e$.

## B.  Stochastic Electron Acceleration by Parallel Right-Handed Circular Polarized Waves

We follow the Hamiltonion method described in Miller & Roberts (1995) to derive the diffusion coefficient for resonant interactions between electrons and transverse, parallel propagating Alfvén and whistler waves. We obtain

$$\tilde{D}_{pp} = \frac{\pi}{2}\, \delta_0^2\, (1 - \mu^2) W_B(k_{\parallel})\, \frac{1}{n_{\parallel}^2}\, \frac{1}{|\beta_g - \beta_{\parallel}|}\,, \quad (B1)$$

where $\mu$ is the pitch-angle cosine, $\delta_0 \equiv m_p/m_e$, $W_B(k_{\parallel})$ is the wave energy density, $n_{\parallel}$ is the index of refraction, $\beta_g$ is the group velocity of the wave, and $\beta_{\parallel} \equiv \beta\mu$. Equation (A1) is written in nondimensional units by dividing $D_{pp}$ by $\Omega_p$. We also write the dimensional wave vector in units of $c/\Omega_p$, so that $k$ is dimensionless throughout this section. Since for Alfvén and Whistler waves the group velocity is typically $\sim \beta_A$, we can approximate $|\beta_g - \beta_{\parallel}| \approx |\beta_{\parallel}|$.

Integrating over pitch angle, we find that

$$\tilde{D}(p) = \frac{\pi}{4}\, \delta_0^2\, W_0\, \frac{1}{\beta} \int_{-1}^{1} d\mu\, \frac{(1 - \mu^2)}{|\mu|}\, k_{\parallel}^{-q}\, \frac{1}{n_{\parallel}^2}\,, \quad (B2)$$

where $W_B = W_0 \, k_\parallel^{-q}$ has been substituted. Using the resonant condition

$$\omega - \beta k_\parallel \mu - \delta_0/\gamma = 0 \tag{B3}$$

and assuming $\omega \ll |k_\parallel \beta \mu|$, $\delta_0/\gamma$, one obtains $\mu \approx -\delta_0/(p \, k_\parallel)$. Thus equation (B2) can be transformed into the expression

$$\tilde{D}(p) = \frac{\pi}{2} \, \delta_0^2 \, W_0 \, \frac{1}{\beta} \int_{k_0}^{k_1} \, dk_\parallel \, k_\parallel^{-q-1} \, (1 - \frac{\delta_0^2}{p^2 k_\parallel^2}) \, \frac{1}{n_\parallel^2} \, , \tag{B4}$$

where $k_1$ is the upper limit on the wavenumber for which waves can still be in resonance with electrons, and $k_1 = \delta_0^{1/2}/\beta_A$ (corresponding to the electron gyrofrequency) while $k_0 = \delta_0/p$, both in normalized units.

For parallel transverse cold plasma waves, the dispersion relation in an e-p plasma can be written as

$$n_\parallel^2 = 1 + \frac{1}{\beta_A^2(\omega + 1)(1 - \omega/\delta_0)} \, , \tag{B5}$$

In Sections 2.2 and 2.3, we took the two limiting forms of equation (B5), namely the dispersion relation for Alfvén waves where

$$n_\parallel^2 = \frac{1}{\beta_A^2} \, , \tag{B6a}$$

with $k_{min} < k_0 < 1/\beta_A$, and the dispersion relation for whistler waves

$$n_\parallel^2 = \frac{1}{\beta_A^4 k_\parallel^2} \, , \tag{B6b}$$

with $1/\beta_A < k_0 < \delta_0^{1/2}/\beta_A$.

The calculated timescales are disjoint at the transition of the two dispersion relations (corresponding to the proton gyrofrequency), as indicated by the dashed lines in Figure 3 which show the fast mode acceleration time scales. Here, instead of using two separate dispersion relations equations (B6a) and (B6b), we use the following dispersion relation which bridges the two regimes:

$$n_\parallel^2 = \frac{1}{\beta_A^2(\beta_A^2 k_\parallel^2 + 1)} \, . \tag{B7}$$

Equation (B7) reduces to equation (B6a) as $k_\parallel \ll 1/\beta_A$ and equation (B6b) as $k_\parallel \gg 1/\beta_A$. In addition, the deviation from equation (B5) when $k_\parallel \approx 1/\beta_A$ can be shown to be small.

Substituting equation (B7) into equation (B4), and deriving the acceleration rate, one finally obtains the timescale for electron reaching Lorentz factor $\gamma$, given by

$$t_E \equiv |\frac{1}{(\gamma - 1)}(\frac{d\gamma}{dt})|^{-1} = \frac{2}{\pi} \, (\frac{q}{q - 1}) \frac{t_{dyn}}{\beta_A^2 \zeta} \, (\frac{r_L}{R})^{2-q} \, (1 - \gamma^{-1}) \, \frac{p^{2-q}}{\beta} \, C(\frac{k_1}{k_0}) \, , \tag{B8}$$

where

$$C\left(\frac{k_1}{k_0}\right) \;=\; \{1 \;+\; \frac{q}{2-q}\delta_0[\left(\frac{k_1}{k_0}\right)^{-q} \;-\; \left(\frac{k_1}{k_0}\right)^{-2}] \;-\; \frac{2}{q}\left(\frac{k_1}{k_0}\right)^{-q}\}^{-1} \;,$$ (B9)

with $k_1/k_0 \;=\; p/(43\beta_A)$. This result is also plotted in Figure 3 by the solid curves.

## Figure Captions

Fig. 1. **(Not shown)** – Model geometry of the system. Magnetic turbulence in the accretion-disk corona surrounding the black hole accelerates particles to high energies through stochastic acceleration.

Fig. 2. – Time scales for proton energy gain and diffusive escape resulting from gyroresonant interactions with parallel propagating Alfvén waves, are shown by the solid curves as a function of dimensionless proton momentum $p_p = \beta_p \gamma_p$. Proton energy-loss time scales through Coulomb interactions and secondary pion production in collisions with particles in the background thermal plasma are shown by the dotted and dot-dashed curves, respectively. The corona is assumed to by 100 gravitational radii in size, and the optical depth and temperature of the background thermal plasma are chosen to be 1 and 51.1 keV (i.e., $\theta_{pl} = 0.1$), respectively. The photo-pair and photo-pion energy-loss time scales are given for dimensionless temperatures $\theta$ of the radiation field equal to $10^{-4}$ and $10^{-1}$ for the long-dashed and short-dashed curves, respectively. All time scales

are plotted in units of the dynamical time scale $t_{dyn}$. The black hole mass is chosen to be $10^8 M_\odot$. (a) Proton energy-loss and energy-gain time scales for a Kolmogorov spectrum ($q = 5/3$). The system is assumed to be accreting at the Eddington luminosity with an equipartition magnetic field. The energy density of plasma turbulence is assumed to be 10% the energy density in the large scale magnetic field. (b) Same parameters as in Fig. 2(a), except that the turbulence spectrum is of the Kraichnan form with $q = 3/2$. (c) Same parameters as in Fig. 2(a), except that the system is assumed to be accreting at 1% of the Eddington luminosity. (d) Same parameters as Fig. 2(a), except that the energy density of plasma turbulence is assumed to be 1% the energy density in the large scale magnetic field, which in turn has an energy density only 1% of its equipartition value. (e) Same parameters as Fig. 2(a), except that black hole mass is $10^9$ M$_\odot$, $\tau_p = 0.01$, and $l_{Edd} = 10^{-4}$.

Fig. 3. – Time scales for electron energy gain and diffusive escape resulting from gyroresonant interactions with fast mode and whistler waves, are shown by the solid curves as a function of dimensionless electron momentum $p_e = \beta_e \gamma_e$. Electron energy-loss time scales through Coulomb and bremsstrahlung interactions with particles in the background thermal plasma are shown by the dashed curves. The standard parameters for the system are the same as given in the caption to Fig. 2a. The Compton energy-loss time scales correspond to dimensionless radiation-field temperatures $\theta = 10^{-4}$ and $10^{-1}$, as indicated on the curves. The synchrotron time scale assumes no self-absorption. All time scales are plotted in units of the $t_{dyn}$. (a) Electron energy-loss and energy-gain time scales for a Kolmogorov spectrum ($q = 5/3$). (b) Same parameters as in Fig. 3(a), except that $q = 3/2$. (c) Same parameters as in Fig. 3(a), except that $l_{Edd} = 0.01$.

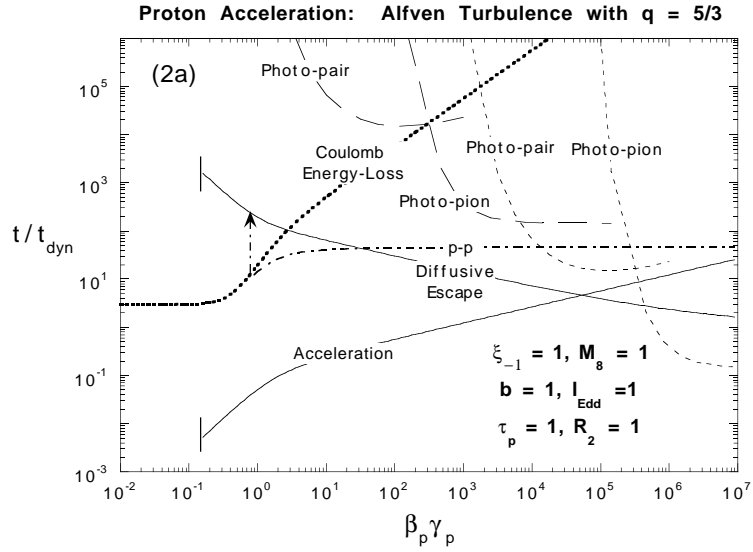

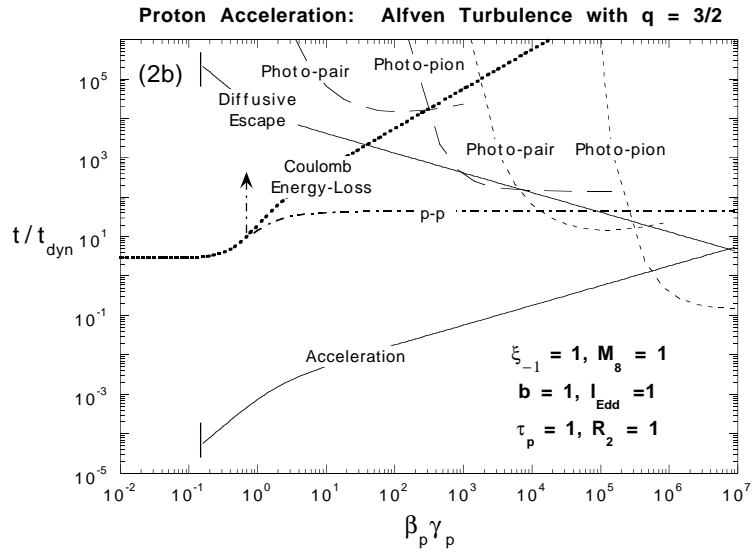

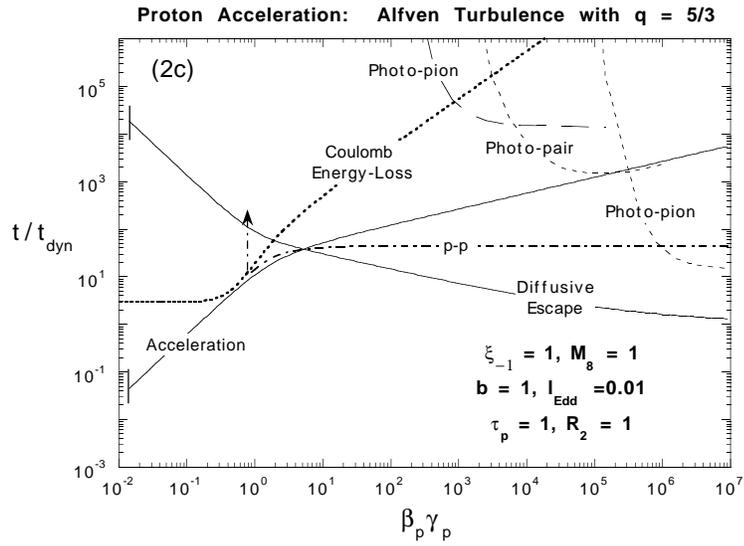

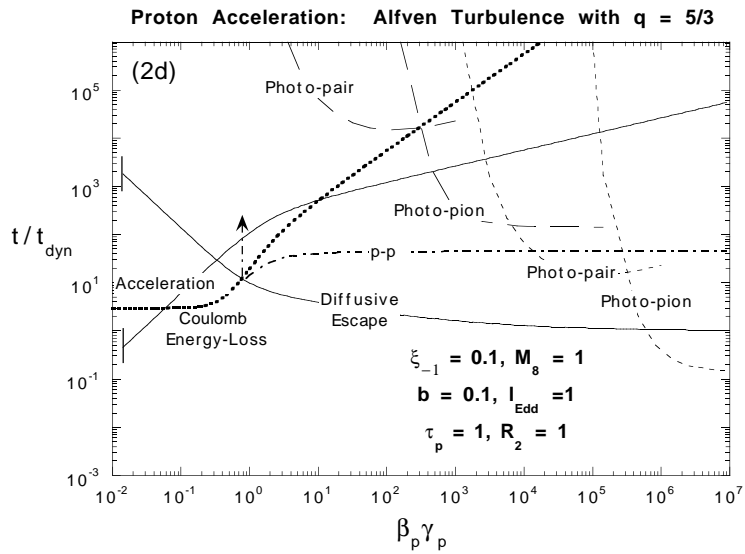

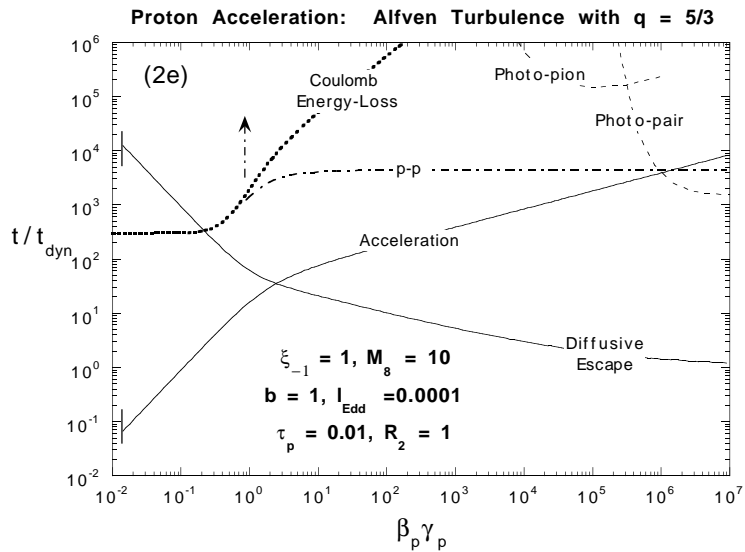

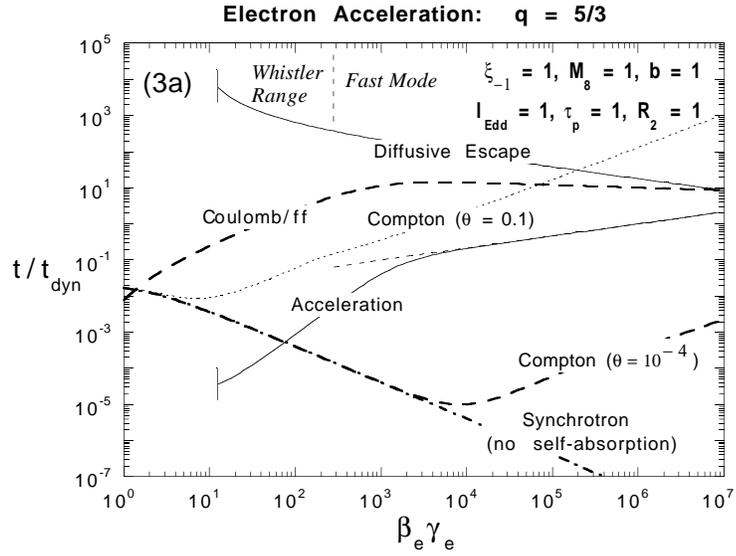

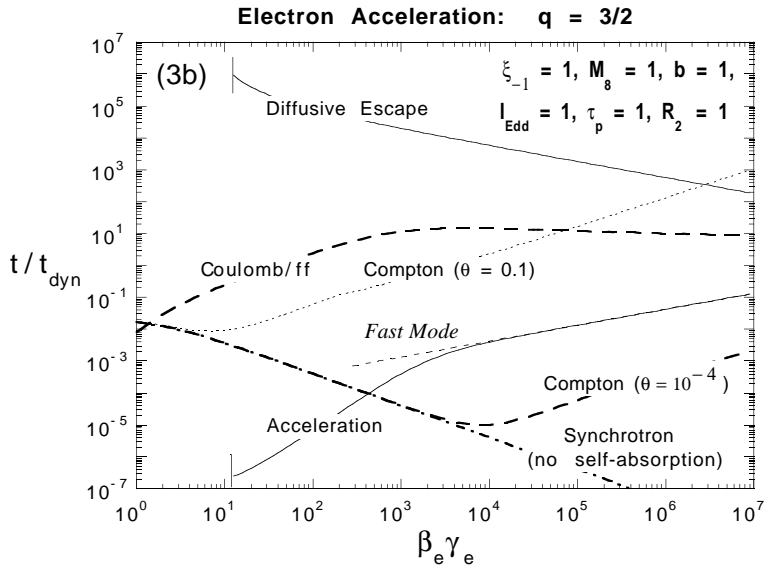

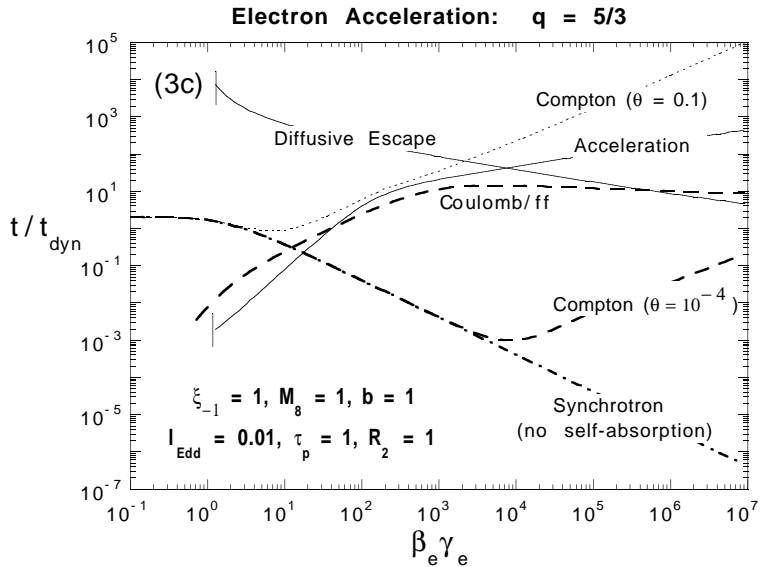